\documentclass[11pt,aps,pra,onecolumn]{revtex4-2}

\usepackage{graphicx}
\usepackage{bm}
\usepackage{enumitem}
\usepackage{physics}
\usepackage{microtype}
\usepackage{xcolor}
\usepackage[colorlinks=true,linkcolor=blue,citecolor=blue,urlcolor=blue]{hyperref}

\begin{document}

\title{Semi-analytical representation of the Gaunt-modified Landau–Lifshitz equation in a plane wave}

\author{S. A. Shekhanov}
\email{sviatoslav.shekhanov@york.ac.uk}
\affiliation{York Plasma Institute, University of York, York, YO10 5DD, UK}

\author{C. P. Ridgers}
\affiliation{York Plasma Institute, University of York, York, YO10 5DD, UK}

\date{\today}

\begin{abstract}
We analyze electron dynamics in a plane electromagnetic wave using the Landau–Lifshitz equation with a quantum radiation reaction correction modeled by a Gaunt factor. In this geometry, the quantum parameter $\chi$ depends solely on the lightfront momentum, allowing the modified equation of motion to retain the integrable structure of the classical problem. We derive a semi-analytical solution for the energy evolution and the four-velocity, which reduces to the known classical result in the appropriate limit. The results provide an analytical and deterministic description of semiclassical radiation reaction in plane-wave fields.
\end{abstract}

\maketitle

\section{Introduction}

Radiation reaction (RR) remains one of the most subtle problems in classical and quantum electrodynamics. 
In interactions between ultra-relativistic electrons and ultra-intense laser fields, radiation emission can substantially modify particle trajectories, leading to strongly nonlinear dynamical effects that are now accessible in experiments at multi-petawatt facilities \cite{DiPiazza2012, MarklundShukla2006, Gonoskov2022}.

The classical theoretical foundation of radiation reaction lies in the Lorentz--Abraham--Dirac equation and is regularized into the Landau--Lifshitz (LL) equation. While LL is formally a perturbative approximation, exact analytical comparisons demonstrate its validity even deep in the ultrarelativistic, radiation-friction-dominated regime \cite{Bulanov2011}. Classical radiation damping in relativistic motion is closely connected to synchrotron radiation theory and its quantum generalization developed in the works of Sokolov and Ternov \cite{SokolovTernov1986}, as well as to early quasiclassical approaches such as the Baier–Katkov formalism \cite{BaierKatkov1998}.

A particularly important configuration for analytical studies is a plane electromagnetic wave. In this geometry the LL equation becomes exactly solvable, as first demonstrated by Di~Piazza \cite{DiPiazza2008}. The integrability originates from the light-front structure of the plane wave: the dynamics reduces to a closed system governed by a single scalar function controlling the evolution of the light-front momentum and, consequently, the particle energy. This exact solution has played a central role in clarifying the structure of classical radiation reaction and serves as a benchmark for numerical implementations.

However, the classical Landau–Lifshitz (LL) equation systematically overestimates radiative losses once the quantum nonlinearity parameter
\begin{equation}\label{eq:chi_parameter}
\chi =
\frac{1}{m E_{cr}}
\sqrt{-(F^{\mu\nu} p_\nu)^2}
\end{equation}
enters the moderately quantum regime, $\chi \sim 0.1$–$1$. In this domain, quantum recoil and the discrete nature of photon emission reduce the average emitted power relative to the classical prediction. The importance of this regime was already emphasized in early strong-field QED analyses and later formalized in the modern theory of nonlinear Compton scattering and pair production \cite{Ritus1985}. Here
\[
E_\text{cr} \equiv \frac{m^2 c^3}{e\hbar} \approx 1.3 \times 10^{16}\,\mathrm{V\,cm^{-1}}
\]
is the Schwinger critical field. The parameter $\chi$ characterizes the field strength in the instantaneous rest frame of the particle in units of $E_\text{cr}$. For an ultra-relativistic particle it can be estimated as $\chi \sim \gamma E_\perp / E_\text{cr}$, where $E_\perp$ is the component of the field transverse to the particle velocity. In a counter-propagating particle–laser geometry, the Lorentz boost enhances the effective field strength in the particle rest frame, leading to the stronger scaling $\chi \sim 2\gamma E_\perp / E_\text{cr}$. As a result, relativistic particles with initial Lorentz factor $\gamma_0 \gg 1$ interacting head-on with sufficiently intense laser pulse can readily reach the regime $\chi \simeq 0.1-1$, where quantum recoil and radiation-reaction effects become non-negligible.

A fully quantum description based on stochastic photon emission provides the most accurate treatment but does not admit closed-form solutions even in simple backgrounds. In practical large-scale simulations, quantum effects are incorporated via Monte Carlo photon emission algorithms embedded in particle-in-cell (PIC) frameworks. This approach, pioneered and developed in works such as Bell \& Kirk \cite{BellKirk2008}, Elkina et al. \cite{Elkina2011}, and Ridgers et al. \cite{Ridgers2014}, forms the basis of modern QED-PIC modeling. Closely related developments by Bulanov and collaborators established the connection between intense laser fields and QED cascade formation in plasma environments \cite{Bulanov2010}, while subsequent analyses by Narozhny and Fedotov clarified fundamental limits of perturbative QED in extreme fields \cite{Narozhny2015,Fedotov2017}.

In contrast, a deterministic semiclassical correction, in which the classical radiation-reaction force is multiplied by a $\chi$-dependent Gaunt factor $g(\chi)$ \cite{Niel2018}, reproduces the quantum-suppressed mean emission rate while preserving the structure of a classical equation of motion, but does not capture the stochastic energy spectrum evolution and broadening associated with discrete emission during radiation cooling \cite{Blackburn2024, Bulanov2024}.

Despite their widespread use in numerical simulations of strong-field QED plasmas, the analytical properties of Gaunt-modified LL equations have received comparatively little attention. In particular, it is not evident whether the integrable structure of the classical plane-wave problem survives once the radiation-reaction force acquires explicit $\chi$-dependence. Clarifying this issue is important both conceptually and practically: plane waves constitute benchmark configurations for testing radiation reaction models, and the existence (or absence) of integrability directly affects the possibility of obtaining exact solutions against which numerical schemes can be validated.

Recent theoretical progress has yielded analytical solutions for quantum radiation reaction by modeling the kinetic evolution of electron beam energy distributions and quantum straggling moments \cite{Blackburn2024, Bulanov2024}. In contrast to ensemble/kinetic treatments that solve Fokker--Planck moment equations for beam energy spread, the present work focuses on single-particle trajectory dynamics governed by the deterministic Gaunt-modified Landau--Lifshitz equation. Specifically, we demonstrate that the exact light-front reduction discovered by Di~Piazza \cite{DiPiazza2008} for classical plane waves remains valid when quantum energy suppression is incorporated via a non-linear Gaunt factor $g(\chi)$. Because $\chi$ in a plane wave depends solely on the light-front momentum $n \cdot u$, the full 4D differential dynamics maps to a 1D scalar Volterra integral equation for $h(\phi)$, enabling an exact closed-form algebraic reconstruction of the four-velocity $u^\mu(\phi)$ and spacetime trajectory $x^\mu(\phi)$ conditional on $h(\phi)$.

We analyze in detail two representative field configurations of direct relevance to laser–plasma interactions:
\begin{enumerate}[label=(\roman*)]
\item a monochromatic plane wave, $a(\phi) = a_0 \sin\phi$, and
\item a finite-duration pulse, $a(\phi) = a_0 \exp\left[-\frac{(\phi - \phi_0)^2}{2(\omega\tau)^2}\right]\sin\phi$.
\end{enumerate}
Here $a_0 \gg 1$ is the dimensionless laser amplitude, $\phi = k \cdot x$ is the laser phase, $\phi_0$ specifies the pulse center phase corresponding to the envelope peak, and $\tau$ characterizes the pulse duration. The normalized amplitude is defined as
\[
a_0 = \frac{1}{mc^2} \frac{e\sqrt{-(F^{\mu\nu} p_\nu)^2}}{k^{\mu}p_{\mu}} = \frac{eE_0}{mc\omega_0},
\]
where $E_0 = |\mathbf{E}| = c|\mathbf{B}|$.

The results provide a fully deterministic and analytical description of semiclassical radiation reaction in a plane wave, thereby bridging the gap between the classical integrable theory and the Gaunt-factor models employed in contemporary numerical simulations.

\section{Model Formulation, Normalization, and Light-Front Reduction}

We consider the motion of an electron of mass $m$ and charge $e$ (here $e > 0$) in an external electromagnetic field described by the field tensor $F^{\mu\nu}(x)$. Throughout this work, we employ natural units $\hbar = c = 1$ and metric signature $(+,-,-,-)$. Greek indices run over spacetime components $0,1,2,3$. We assume sums over repeating indices. The four-dimensional product of two arbitrary four-vectors $a_\mu$ and $b_\mu$ is indicated as $a \cdot b$, i.e., $a\cdot b = a_\mu b^\mu$.

\subsection{Geometry, Unit Conventions, and Parameter Normalization}

A plane wave propagating along direction $\mathbf{n}$ is characterized by the wave four-vector $k^\mu = \omega_0 n^\mu$, where $n^\mu = (1, \mathbf{n})$ is a light-like vector ($n^2 = 0$) and $\omega_0$ is the characteristic wave frequency. The dimensionless wave phase is defined as
\begin{equation}
\phi = k \cdot x = \omega_0 (t - \mathbf{n} \cdot \mathbf{x}).
\label{phi_definition}
\end{equation}

In what follows, we adopt normalized units in which the central frequency is set to unity ($\omega_0 = 1$), so that the wave vector simplifies to $k^\mu \to n^\mu$ and the phase reduces to $\phi = n \cdot x$. This choice omits redundant factors of $\omega_0$ throughout the analytical derivations without loss of generality. Dimensional physical quantities for an optical laser ($\lambda_0 = 1\,\mu\mathrm{m}$) are explicitly restored in Table~\ref{tab:units}.

The four-potential of an arbitrary plane wave is given by
\begin{equation}
A^\mu(\phi) = \sum_j a_j^\mu \psi_j(\phi) \quad (j=1,2),
\label{A_plane_wave}
\end{equation}
where $a_j^\mu$ are constant polarization four-vectors satisfying transversality $n \cdot a_j = 0$ and orthogonality $a_i \cdot a_j = - a_0^2 (m/e)^2 \delta_{ij}$, and $\psi_j(\phi)$ are dimensionless envelope functions. Taking $F^{\mu\nu} = \partial^\mu A^\nu - \partial^\nu A^\mu = n^\mu A'^\nu - n^\nu A'^\mu$, the field tensor takes the concise form
\begin{equation}
F^{\mu\nu}(\phi) = \sum_j f_j^{\mu \nu}\psi'_j(\phi), \qquad f_j^{\mu \nu} = n^\mu a_j^\nu - n^\nu a_j^\mu,
\label{F_plane_wave}
\end{equation}
where primes denote differentiation with respect to the phase $\phi$. We define the normalized vector potentials $\xi_j^\mu = e a_j^\mu / m$, such that $\sum_j (\xi_j \psi'_j)^2 = a_0^2 \psi'^2(\phi)$ characterizes the local field amplitude.

\begin{table}[h!]
\caption{\label{tab:units}
Summary of physical variables, mathematical definitions, and characteristic values for an optical laser pulse ($\lambda_0 = 1\,\mu\mathrm{m}$) in a counter-propagating electron configuration in natural units ($\hbar = c = 1$).}
\begin{ruledtabular}
\begin{tabular}{llll}
Quantity & Symbol & Definition / Expression & Characteristic Value \\
\hline
Laser wavelength & $\lambda_0$ & $2\pi / \omega_0$ & $1\,\mu\mathrm{m}$ \\
Laser frequency & $\omega_0$ & $2\pi / \lambda_0$ & $1.88\times10^{15}\,\mathrm{s}^{-1}$ \\
Wave four-vector & $k^\mu$ & $k^\mu = \omega_0 n^\mu$ & $\mathrm{s}^{-1}$ \\
Null direction vector & $n^\mu$ & $(1,0,0,1),\quad n^2=0$ & Dimensionless \\
Laser phase & $\phi$ & $k\cdot x = \omega_0 n\cdot x$ & Dimensionless \\
Normalized pulse duration & $\omega\tau$ & $\omega_0 \tau$ & $10$ \\ 
Pulse center phase & $\phi_0$ & $5\omega\tau$ & $50$ \\ 
Mass-to-frequency ratio & $a_S$ & $m / \omega_0 = 1/\eta$ & $4.12\times10^{5}$ \\
Normalized laser amplitude & $a_0$ & $e\sqrt{-a^2}/m$ & $50$ \\
Polarization vectors & $\xi_j^\mu$ & $\xi_j^\mu=e a_j^\mu/m,\quad n\cdot\xi_j=0,\quad \xi_i\cdot\xi_j=-a_0^2\delta_{ij}$ & Dimensionless \\
Initial light-front factor & $\rho_0$ & $n\cdot u_0 = \gamma_0(1-\beta_{z,0})$ & $\rho_0 \simeq 2\gamma_0 = 30000$ \\
Radiation reaction parameter & $\tau_R$ & $\frac{2}{3}\alpha\eta = \frac{2e^2\omega_0}{3m}$ & $1.18\times10^{-8}$ \\
Peak quantum parameter & $\chi_{\mathrm{peak}}$ & $\frac{\rho_0 a_0}{a_S} = \eta \rho_0 a_0$ & $3.7$ \\
Initial electron energy & $\mathcal{E}_0$ & $\gamma_0 m$ & $7.67\,\mathrm{GeV}$ ($\gamma_0 = 15000$) \\
\end{tabular}
\end{ruledtabular}
\end{table}

\subsection{Semiclassical Landau--Lifshitz Framework and Model Validity}

The classical Landau--Lifshitz (LL) equation \cite{LandauLifshitz} reads
\begin{equation}
m\frac{d u^\mu}{ds}
=
e F^{\mu\nu} u_\nu
+
\tau_R
\left[
e \partial_\alpha F^{\mu\nu} u^\alpha u_\nu
-
\frac{e^2}{m} F^{\mu\nu} F_{\alpha\nu} u^\alpha
+
\frac{e^2}{m}
(F^{\alpha\beta} u_\beta F_{\alpha\gamma} u^\gamma)
u^\mu
\right],
\label{LL_classical}
\end{equation}
where
\begin{equation}
\tau_R = \frac{2 e^2}{3 m}
\end{equation}
is the classical radiation-reaction parameter (in units where $\omega_0=1$, see Table \ref{tab:units}).

To account for the quantum suppression of radiation emission, we follow the semiclassical prescription in which the radiation-friction force is scaled by the quantum Gaunt factor $g(\chi) \le 1$\cite{BaierKatkovStrakhovenko1974}. In the classical Landau–Lifshitz equation, the quadratic-field $O(F^2)$ terms describe irreversible energy-momentum loss due to photon emission, whereas the field-derivative term accounts for classical field-gradient (Schott) effects. Because quantum photon recoil acts specifically on the radiative emission process, $g(\chi)$ scales the paired radiation-friction vector:

\begin{equation}
m\frac{d u^\mu}{ds}
=
e F^{\mu\nu} u_\nu
+
\tau_R
\left\{
e \partial_\alpha F^{\mu\nu} u^\alpha u_\nu
+
g(\chi)\left[
-
\frac{e^2}{m} F^{\mu\nu} F_{\alpha\nu} u^\alpha
+
\frac{e^2}{m}
(F^{\alpha\beta} u_\beta F_{\alpha\gamma} u^\gamma)
u^\mu
\right]
\right\}.
\label{LL_gaunt_tau}
\end{equation}

In continuous semiclassical PIC and kinetic solvers (such as SMILEI), field gradients on sub-wavelength scales are neglected, causing the field-derivative term to drop out and reducing Eq.~\eqref{LL_gaunt_tau} directly to the quantum-corrected Landau--Lifshitz (cLL) model evaluated in continuous radiation-reaction routines \cite{Niel2018PPCF, Niel2018}.

By scaling the paired quadratic terms together, this formulation preserves the four-velocity normalization condition $u^\mu u_\mu = 1$ (i.e., $u_\mu \, du^\mu/ds = 0$) identically for any $g(\chi)$, while leaving the classical field-gradient term unscaled. In the ultrarelativistic regime ($\gamma \gg 1$), the $O(F^2)$ friction terms dominate the radiative cooling dynamics, ensuring that the total radiated power correctly reduces to $I_{\mathrm{rad}} = g(\chi) I_{\mathrm{cl}}$.

For a plane wave, this prescription has an additional useful property. Upon contraction of Eq.~\eqref{LL_gaunt_tau} with the wave vector $n_\mu$, the field-gradient contribution vanishes because of plane-wave transversality,
$F^{\mu\nu}n_\mu=0$. Consequently, the corresponding light-front evolution retains the simple form associated with the dominant radiation-reaction contribution.

The quantum suppression factor $g(\chi)$ is defined as the ratio of the full quantum radiation intensity to the intensity emitted by a classical electron. Following \cite{Bulanov2024}, it can be expressed in terms of the Airy function as
\begin{equation}
\label{eq:gaunt_exact}
g(\chi)
=
-\frac{3}{4}
\int_0^\infty
\frac{
4+5\chi x^{3/2}+4\chi^2 x^3
}{
\left(1+\chi x^{3/2}\right)^4
}
\operatorname{Ai}'(x)\,x\,dx .
\end{equation}
In the classical limit ($\chi\ll1$), this expression approaches unity according to
\begin{equation}
g(\chi)
=
1-\frac{55\sqrt{3}}{16}\chi+\ldots
\approx
1-5.9\chi+\ldots .
\end{equation}

For numerical calculations, we employ the continuous fitting function given in \cite{Esirkepov2015},
\begin{equation}
\label{eq:gaunt_fit}
g(\chi)
=
\left[
1+18\chi+69\chi^2+73\chi^3+5.806\chi^4
\right]^{-1/3}.
\end{equation}
This fit reproduces the full radiation-intensity result to within $1\%$ over $0<\chi<20$ and correctly recovers the strong-field scaling $g(\chi)\propto\chi^{-4/3}$ at large $\chi$. Thus, $g(\chi)$ accounts for the quantum suppression of average radiative losses due to electron recoil.

\begin{figure}[h!]
   \centering
   \includegraphics[width=0.48\linewidth]{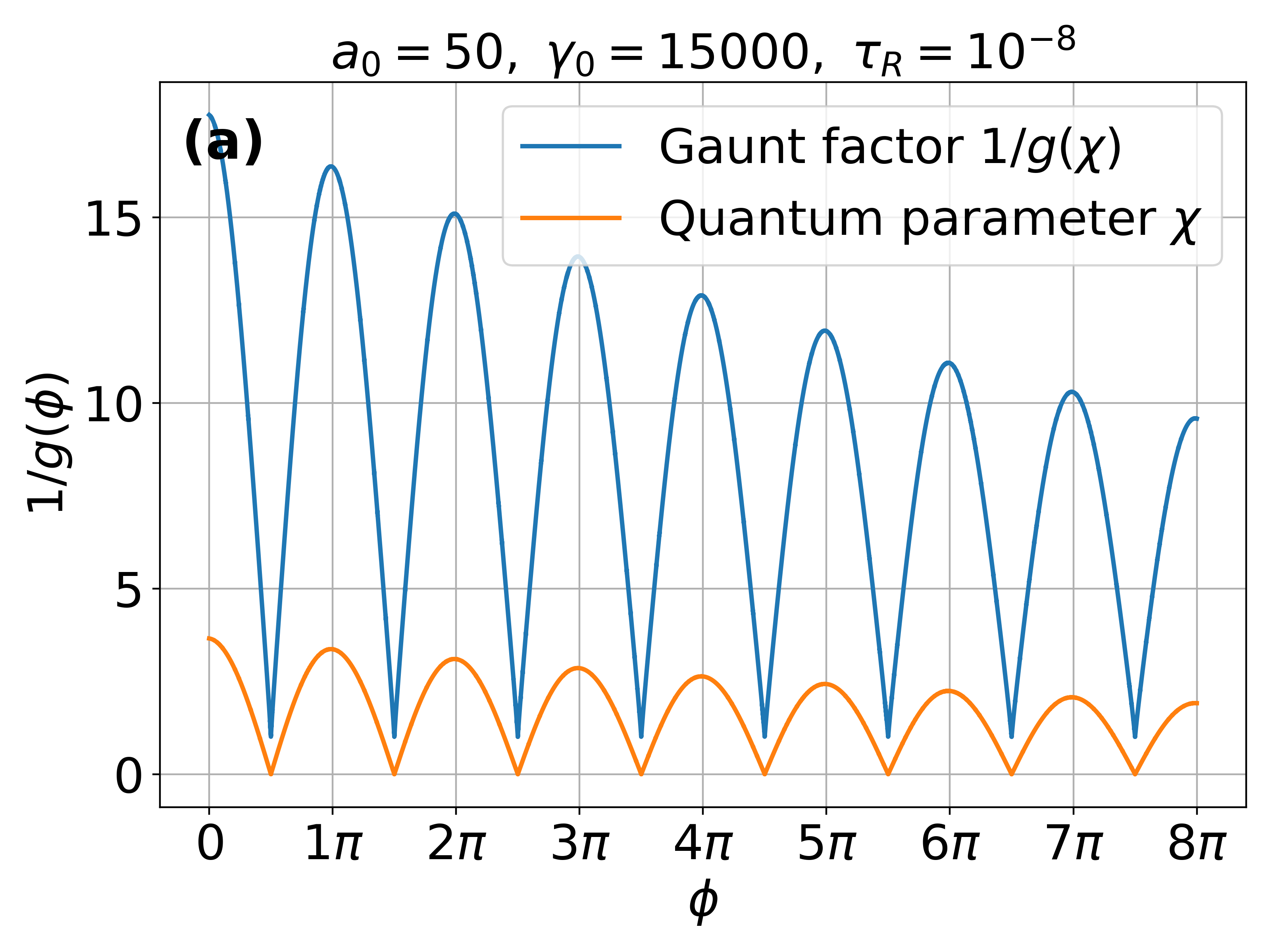}
   \includegraphics[width=0.48\linewidth]{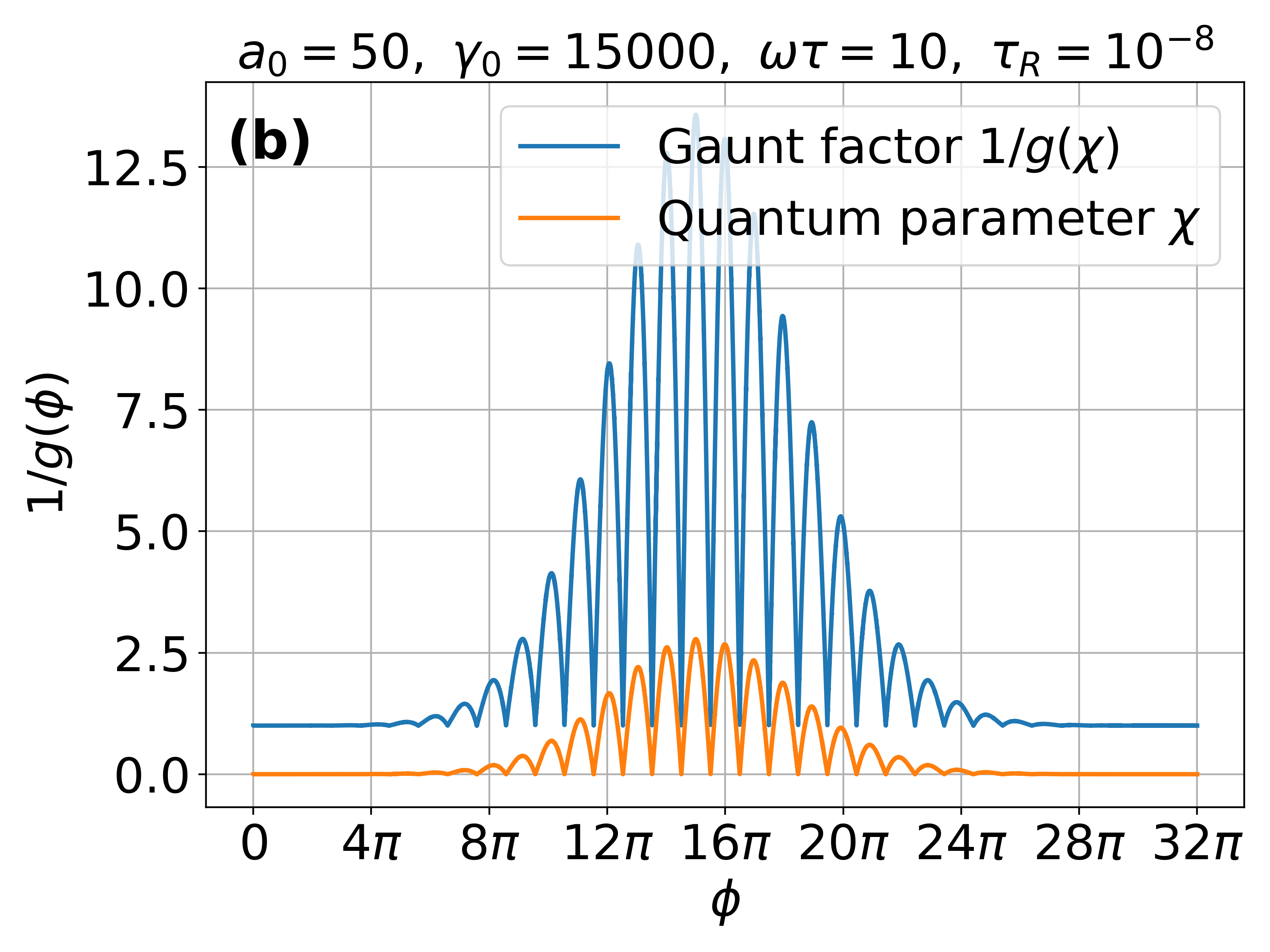}
   \caption{Phase dependence of the inverse Gaunt factor $1/g(\chi)$ (blue) and quantum parameter $\chi$ (orange) in (a) a monochromatic plane wave and (b) a Gaussian-enveloped pulse with $a_0=50$, $\gamma_0=1.5 \times 10^4$, $\omega\tau=10$, $\phi_0 = 5\omega\tau$ and $\tau_R=10^{-8}$. Case of a counter-propagating relativistic particle.}
   \label{fig:gaunt}
\end{figure}

The evolution of $g(\chi(\phi))^{-1}$ along the trajectory is shown in Fig.~\ref{fig:gaunt}. Since the minima of $g(\chi)$ correspond to the maxima of $\chi$, the resulting dynamics exhibits a periodic modulation synchronized with the laser phase. For a monochromatic wave [Fig.~\ref{fig:gaunt}(a)], this produces a slowly decaying oscillatory pattern due to radiative energy loss ($h \propto \chi^{-1}$). For a Gaussian pulse [Fig.~\ref{fig:gaunt}(b)], $g(\chi)^{-1}$ instead mirrors the pulse envelope, rising to a maximum ($\chi \approx 2.5$) near the center before decaying. Consequently, the strong interaction region for both $g(\chi)^{-1}$ and $h(\phi)$ is localized to $\phi \in [8\pi, 24\pi]$, outside of which $h(\phi)$ plateaus as the particle leaves the field.

\subsection{Phase Formulation and Decoupling of Light-Front Momentum}

Since $F^{\mu\nu}$ depends only on $\phi$,
it is advantageous to use $\phi$ as evolution parameter.
Using
\[
\frac{d}{ds} = (n\cdot u)\frac{d}{d\phi},
\]
Eq.~(\ref{LL_gaunt_tau}) becomes
\begin{equation}
m\frac{d u^\mu}{d\phi}
=
\frac{1}{n\cdot u}
\left\{
e F^{\mu\nu} u_\nu
+
\tau_R
\left[
e \partial_\alpha F^{\mu\nu} u^\alpha u_\nu
+
g(\chi)\left(
-\frac{e^2}{m} F^{\mu\nu} F_{\alpha\nu} u^\alpha
+
\frac{e^2}{m}
(F^{\alpha\beta} u_\beta F_{\alpha\gamma} u^\gamma)
u^\mu
\right)
\right]
\right\}.
\label{LL_gaunt_phi}
\end{equation}

We now introduce the light-front momentum
\begin{equation}
\rho(\phi) = n \cdot u(\phi).
\label{rho_def}
\end{equation}

Using the plane-wave identities
\begin{align}
F^{\mu\nu} n_\nu &= 0, \\
F^{\mu\nu}F_{\nu\alpha}
&=
- (a_j\psi'_j)^2 n^\mu n_\alpha,
\end{align}
one finds
\begin{equation}
(F^{\alpha\beta}u_\beta)
(F_{\alpha\gamma}u^\gamma)
=
- \rho^2 (a_j\psi'_j)^2.
\end{equation}

Contracting Eq.~(\ref{LL_gaunt_phi}) with $n_\mu$, all Lorentz-force terms vanish due to transversality,
and the equation reduces to a scalar evolution equation:
\begin{equation}
\frac{d}{d\phi}\left(\frac{1}{\rho}\right)
=
\tau_R g(\chi)
\sum_j\xi_j^2\psi^{'2}_j.
\label{rho_equation}
\end{equation}

This equation is exact and fully decoupled from transverse components. It is the central structural property responsible for integrability.

\subsection{Dissipation Function \texorpdfstring{$h(\phi)$}{h(phi)} and Scalar Volterra Quadrature}

Following the classical light-front reduction, we define the scalar light-front dissipation function $h(\phi)$ via
\begin{equation}
\rho(\phi) \equiv n\cdot u(\phi) = \frac{\rho_0}{h(\phi)},
\qquad
\rho_0 = n\cdot u(\phi_0).
\label{h_def}
\end{equation}
Substituting Eq.~(\ref{h_def}) into the scalar light-front equation yields the initial-value problem for $h(\phi)$:
\begin{equation}\label{eq:h_function}
\frac{dh}{d\phi}
=
\tau_R \rho_0 g(\chi)\sum_j\xi_j^2\psi^{'2}_j(\phi), \qquad h(\phi_0) = 1.
\end{equation}
In a plane wave, the local quantum nonlinearity parameter $\chi$ depends nonlinearly on $h(\phi)$:
\begin{equation}\label{eq:chi}
\chi(\phi)
=
\frac{\rho_0}{m\,h(\phi)}
\sqrt{\sum_j\xi_j^2\psi^{'2}_j(\phi)}.
\end{equation}
Integrating Eq.~(\ref{eq:h_function}) yields the scalar non-linear Volterra integral equation
\begin{equation}
h(\phi)
=
1 + 
\tau_R \rho_0
\int_{\phi_0}^{\phi}d\varphi\,
g\Big[\chi(\varphi, h(\varphi))\Big]\sum_j\xi_j^2\psi^{'2}_j(\varphi).
\label{h_quadrature}
\end{equation}
Eq.~(\ref{h_quadrature}) represents an exact scalar reduction of the full 4D differential dynamics. Once $h(\phi)$ is determined, either analytically in asymptotic regimes or through 1D scalar numerical integration, all components of the four-velocity $u^\mu(\phi)$ can be reconstructed algebraically.

\subsubsection{Analytical Solutions in Asymptotic Limits}
While Eq.~(\ref{h_quadrature}) cannot be integrated in closed elementary form for arbitrary pulse shapes across all $\chi$, $h(\phi)$ is analytically solvable in key physical limits:

1. \textbf{Ultra-Quantum Limit ($\chi \gg 1$):} In the strong-field QED regime ($\chi \gg 1$), the ratio of QED to classical radiated power scales as $g(\chi) \approx c_0 \chi^{-4/3}$ (where $c_0 \approx 0.556$). Substituting $\chi(\phi)$ from Eq.~(\ref{eq:chi}) into Eq.~(\ref{eq:h_function}) yields:
\begin{equation}
\frac{dh}{d\phi} = c_0 \tau_R \rho_0 \left( \frac{m}{\rho_0} \right)^{4/3} \left( \sum_j \xi_j^2 \psi_j'^2(\phi) \right)^{1/3} h(\phi)^{4/3}.
\end{equation}
Separating variables and integrating directly provides an exact closed-form solution for \emph{any} laser envelope profile:
\begin{equation}
h(\phi) = \left[ 1 - \frac{1}{3} c_0 \tau_R \rho_0 \left( \frac{m}{\rho_0} \right)^{4/3} \int_{\phi_0}^{\phi} d\varphi \left( \sum_j \xi_j^2 \psi_j'^2(\varphi) \right)^{1/3} \right]^{-3}.
\end{equation}

2. \textbf{Constant Amplitude Field ($\sum_j \xi_j^2 \psi_j'^2 = a_0'^2 = \text{const}$):} For a constant field envelope, Eq.~(\ref{eq:h_function}) becomes autonomous, $\frac{dh}{d\phi} = \tau_R \rho_0 a_0'^2 g\left(\frac{\chi_0}{h}\right)$ with $\chi_0 = \frac{\rho_0 a_0'}{m}$. Variable separation provides an exact implicit analytical solution $\phi(h)$ for any choice of $g(\chi)$:
\begin{equation}
\int_{1}^{h(\phi)} \frac{dh'}{g\left(\frac{\chi_0}{h'}\right)} = \tau_R \rho_0 a_0'^2 (\phi - \phi_0).
\end{equation}

\subsubsection{Numerical Implementation and Scheme Validation}
For general field envelopes (e.g., finite Gaussian or flat-top sinusoidal profiles), Eq.~(\ref{eq:h_function}) is integrated as a 1D scalar initial-value problem using an adaptive 8th-order Dormand--Prince integrator (\texttt{DOP853}) with relative and absolute tolerances set to $10^{-11}$ and $10^{-13}$, respectively.

The numerical integration specifies an initial light-front phase $\phi_0 = 0$, initial electron Lorentz factor $\gamma_0 = 1.5 \times 10^4$, counter-propagating geometry ($n \cdot u(\phi_0) = \gamma_0 + \sqrt{\gamma_0^2 - 1} \approx 2\gamma_0$), polarization vectors $\xi_j$, field derivatives $\psi_j'(\phi)$, and dimensionless parameters $a_0 = 50$, $\omega\tau = 10$, and $\tau_R = 10^{-8}$.

Numerical precision and physical self-consistency are continuously verified against the following key invariants:
\begin{enumerate}
    \item \textbf{Four-velocity normalization constraint:} $|u^\mu u_\mu - 1| \lesssim 10^{-7}$. For ultra-relativistic particles ($\gamma_0 = 1.5 \times 10^4$), $(u^0)^2 \approx 2.25 \times 10^8$. Subtracting components of this magnitude in 64-bit floating-point arithmetic ($\epsilon_{\text{mach}} \approx 2.2 \times 10^{-16}$) imposes a fundamental roundoff cancellation floor of $\mathcal{O}(\gamma_0^2 \epsilon_{\text{mach}}) \approx 5 \times 10^{-8}$. Thus, the analytical reconstruction remains strictly on-shell down to machine precision limits.
    \item \textbf{Light-front momentum conservation:} $|n\cdot u(\phi) - \rho_0/h(\phi)| < 10^{-11}$, bounded directly by the scalar ODE integration precision.
\end{enumerate}

\begin{figure}
    \centering
    \includegraphics[width=0.48\linewidth]{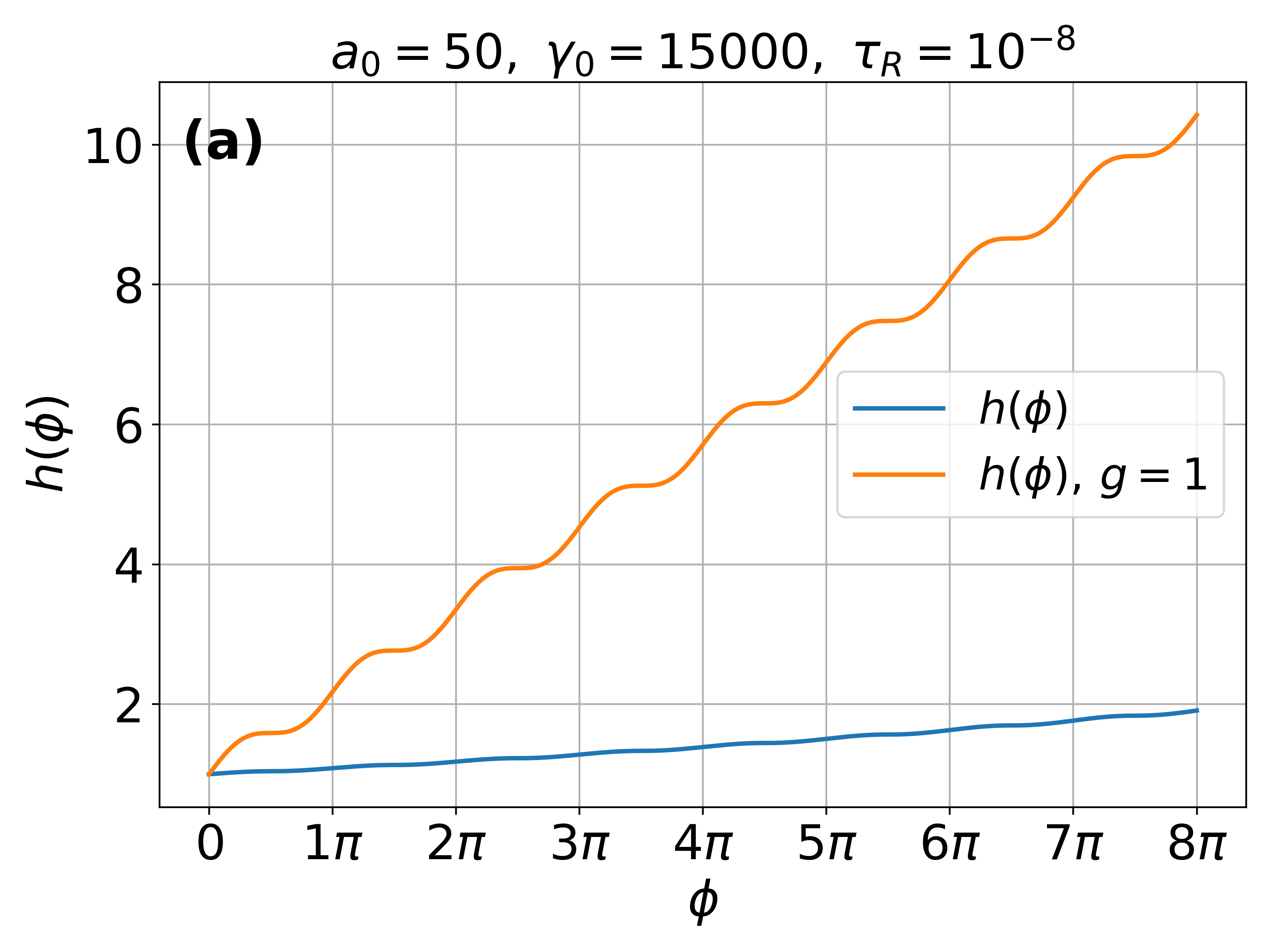}
    \includegraphics[width=0.48\linewidth]{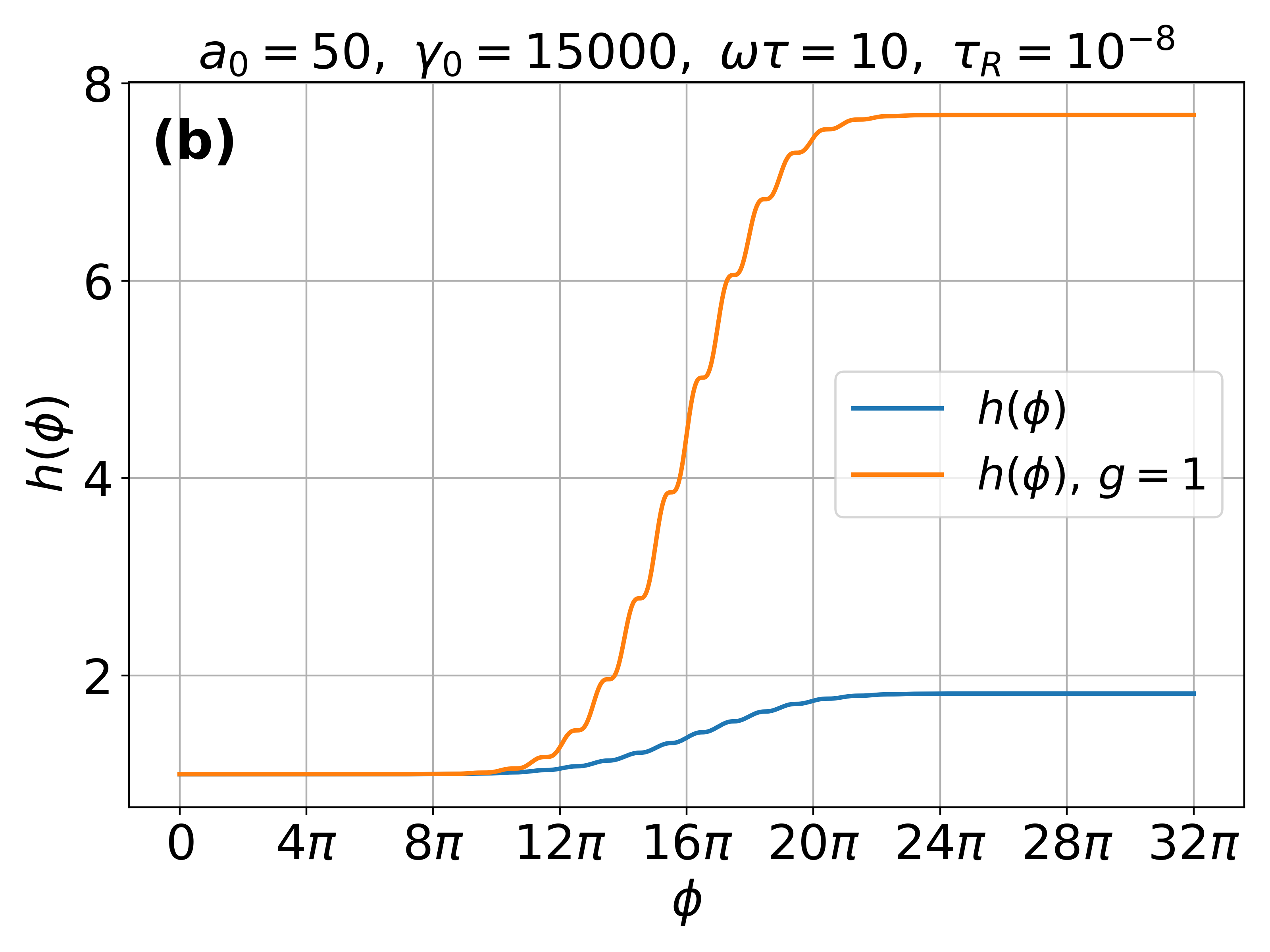}
    \caption{Evolution of the light‑front dissipation function $h(\phi)$ in the classical radiation-reaction (RR) limit ($g=1$, orange) and the semiclassical Gaunt-modified regime (blue) for (a) a monochromatic wave and (b) a finite‑duration Gaussian pulse ($a_0=50$, $\gamma_0 = 1.5 \times 10^4$, $\omega\tau = 10$, $\phi_0 = 5\omega\tau$, $\tau_R = 10^{-8}$) in a counter-propagating geometry.}
    \label{fig:h_function}
\end{figure}

Figure~\ref{fig:h_function} compares the classical and Gaunt-modified dynamics, illustrating quantum suppression of energy loss at large $\chi$. Physically, $h(\phi)$ directly governs the light-front momentum projection $n\cdot u(\phi)$, which sequentially mediates the remaining momentum components. 

For a monochromatic wave [Fig.~\ref{fig:h_function}(a)], the continuous energy drainage per cycle produces an unbounded secular increase in $h(\phi)$. In contrast, for a finite Gaussian pulse [Fig.~\ref{fig:h_function}(b)], $h(\phi)$ grows only during the pulse overlap region and saturates to a constant asymptotic plateau post-interaction.

This dynamic is further detailed in Fig.~\ref{fig:four_velocity}. Under pure Lorentz-force dynamics (the classical Lorentz baseline), the instantaneous energy $u^0(\phi)$ undergoes intense oscillations inside the wave field; however, for a finite pulse, it yields no net asymptotic energy transfer, fully consistent with the assumptions and boundary conditions of the Lawson–Woodward theorem. When radiation reaction is included, time-reversal symmetry is broken: fast field-driven oscillations remain superimposed on the electron energy profile, but accompany an overall monotonic secular decrease, leading to an asymptotic energy deficiency and a net violation of Lawson–Woodward conditions.

\section{Reconstruction of Four-Velocity, Trajectory, and Numerical Verification}

\subsection{Closed-Form Reconstruction of \texorpdfstring{$u^\mu(\phi)$}{u^mu(phi)} and Trajectory \texorpdfstring{$x^\mu(\phi)$}{x^mu(phi)}}

The reduced four-velocity $\tilde u^\mu(\phi) \equiv h(\phi) u^\mu(\phi)$ satisfies a linear matrix differential equation in a plane-wave background. Due to the algebraic closure and light-like properties of plane-wave field tensors,
\begin{equation}
f_i^{\mu\lambda} {f_j}_\lambda{}^\nu n_\nu = 0, \qquad f^{\mu\lambda} f_{\lambda\sigma} f^{\sigma\nu} = 0,
\end{equation}
the matrix exponential (time-ordered Dyson series) truncates exactingly at second order.

The exact four-velocity is thus reconstructed in closed form as
\begin{equation}
u^\mu(\phi)=\frac{1}{h(\phi)}
\Bigl\{
u_0^\mu
+
\frac{h^2(\phi)-1}{2\rho_0}n^\mu
+
\frac{1}{\rho_0}\sum_j\mathcal{I}_j(\phi)
\frac{ef_j^{\mu\nu}}{m}u_{0,\nu}
+
\frac{1}{2\rho_0}
\sum_j\xi_j^2\mathcal{I}_j^2(\phi) n^\mu
\Bigr\},
\label{eq:four_velocity_closed}
\end{equation}
where the phase-dependent scattering integrals are defined by
\begin{equation}
\mathcal{I}_j(\phi)
=
\int_{\phi_0}^{\phi}
\left[
h(\varphi)\psi'_j(\varphi)
+
\tau_R \rho_0\psi''_j(\varphi)
\right] d\varphi.
\end{equation}

Physically, Eq.~\eqref{eq:four_velocity_closed} reveals that radiation reaction preserves the canonical structure of free Lorentz motion in a plane wave. The damping and radiation friction effectively dress the normalized vector potential via the phase integrals $\mathcal{I}_j(\phi)$, while the global $1/h(\phi)$ factor accounts for the cumulative light-cone energy loss.

To obtain the particle trajectory $x^\mu(\phi)$, we note that the light-front momentum relation $n \cdot u = \rho_0/h(\phi)$ yields the differential phase mapping $d\tau/d\phi = h(\phi)/\rho_0$. Consequently, the explicit $1/h(\phi)$ damping factor cancels in the coordinate derivative:
\begin{equation}
\frac{dx^\mu}{d\phi} = u^\mu(\phi) \frac{d\tau}{d\phi} = \frac{\tilde{u}^\mu(\phi)}{\rho_0}.
\end{equation}

Direct integration from initial position $x_0^\mu = x^\mu(\phi_0)$ yields the closed-form spacetime trajectory:
\begin{equation}
x^\mu(\phi) = x_0^\mu + \frac{\phi - \phi_0}{\rho_0} u_0^\mu + \frac{\mathcal{J}_0(\phi)}{2\rho_0^2} n^\mu + \frac{1}{\rho_0^2} \sum_j \mathcal{J}_{1,j}(\phi) \frac{e f_j^{\mu\nu}}{m} u_{0,\nu} + \frac{1}{2\rho_0^2} \sum_j \xi_j^2 \mathcal{J}_{2,j}(\phi) n^\mu,
\end{equation}
where the nested coordinate phase integrals are defined as
\begin{equation}
\mathcal{J}_0(\phi) = \int_{\phi_0}^{\phi} \left[h^2(\varphi) - 1\right] d\varphi, \quad 
\mathcal{J}_{1,j}(\phi) = \int_{\phi_0}^{\phi} \mathcal{I}_j(\varphi) \, d\varphi, \quad 
\mathcal{J}_{2,j}(\phi) = \int_{\phi_0}^{\phi} \mathcal{I}_j^2(\varphi) \, d\varphi.
\end{equation}

When quantum radiation-reaction effects are turned off [e.g., setting $g(\chi) \to 1$], Eq.~\eqref{eq:four_velocity_closed} reduces to the classical Di Piazza solution.

\subsection{Particle Dynamics and Numerical Model Validation}

\begin{figure}
    \centering
    \includegraphics[width=0.48\linewidth]{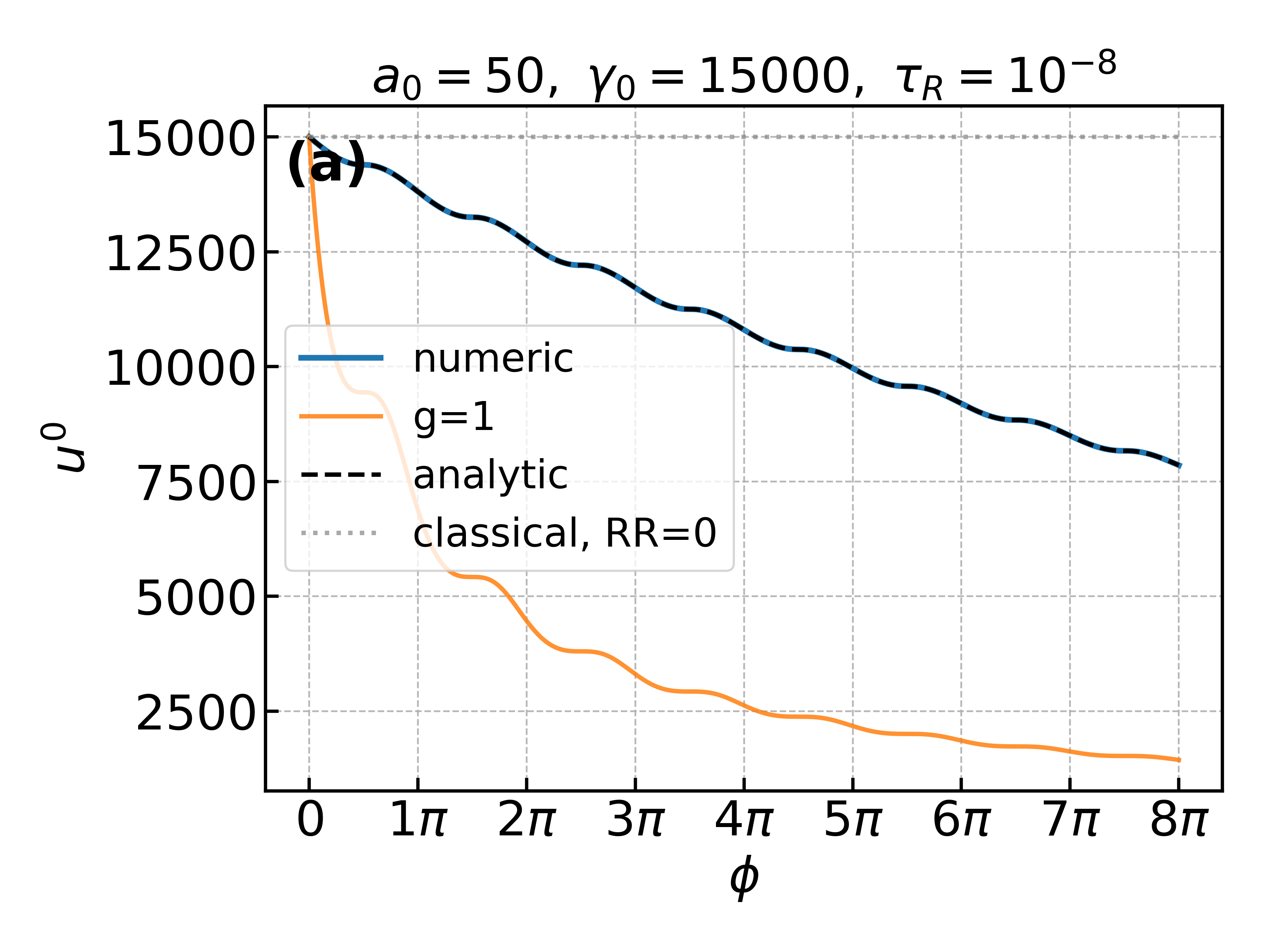}
    \includegraphics[width=0.48\linewidth]{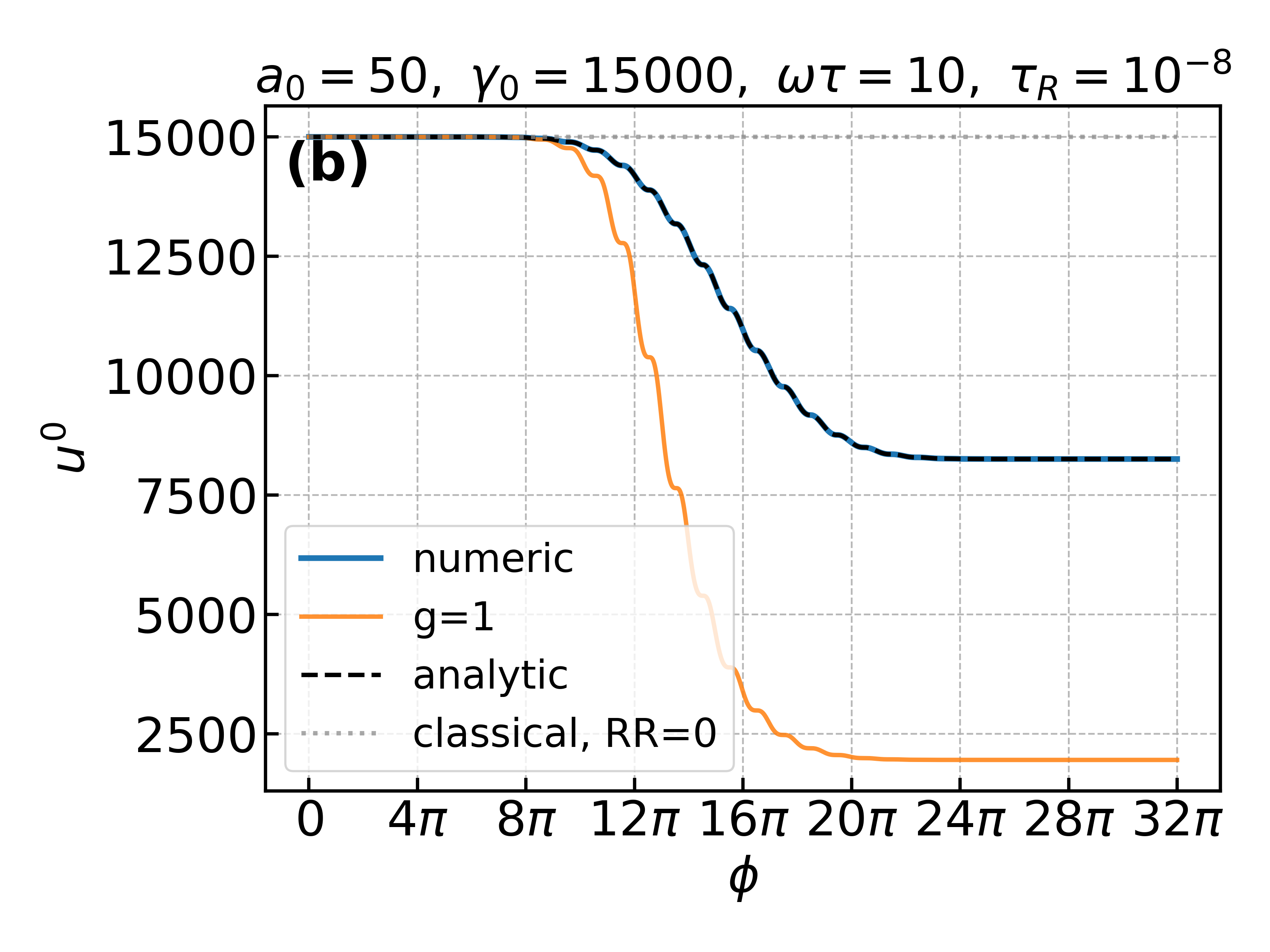}\\
    \includegraphics[width=0.48\linewidth]{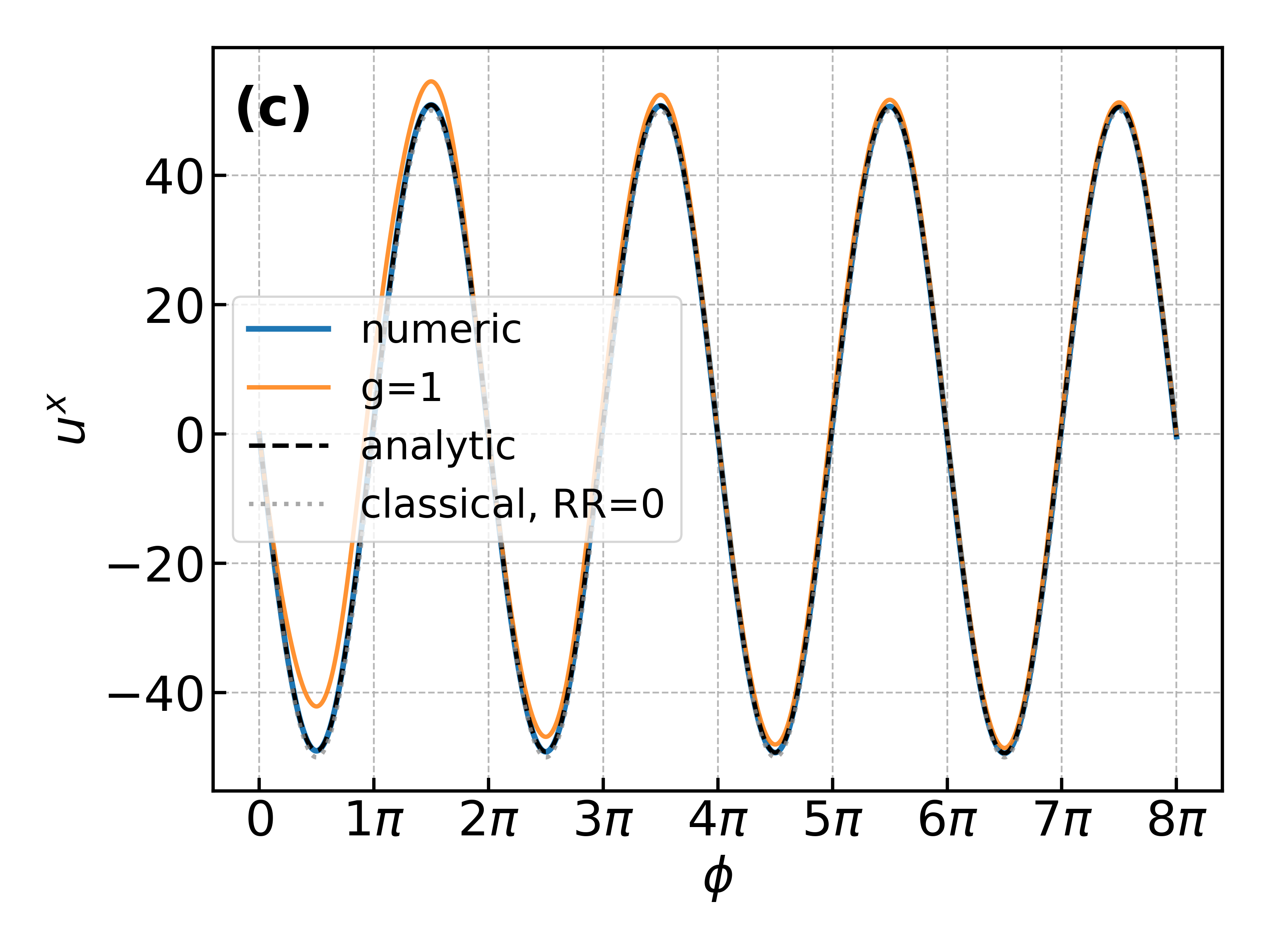}
    \includegraphics[width=0.48\linewidth]{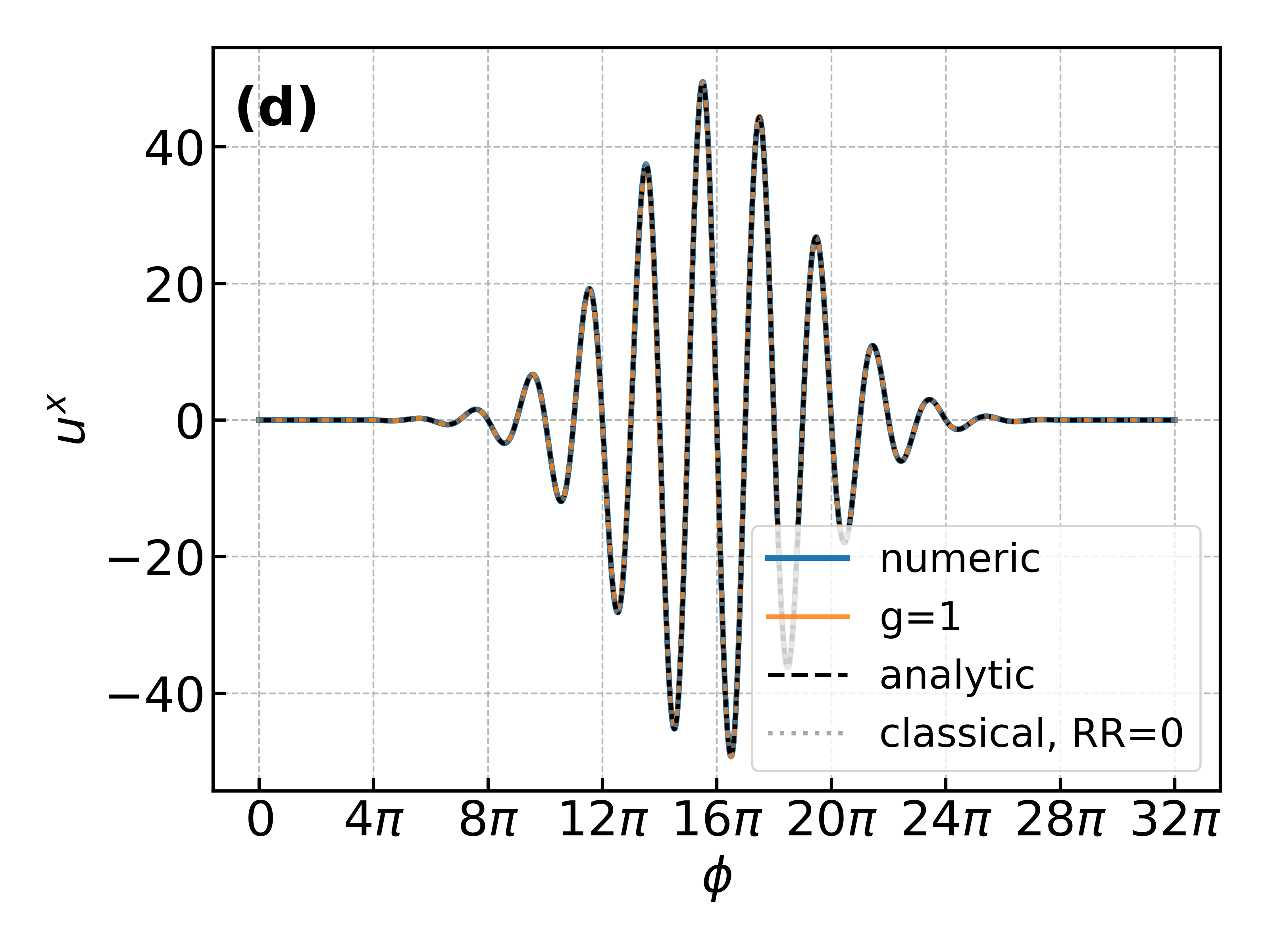}\\
    \includegraphics[width=0.48\linewidth]{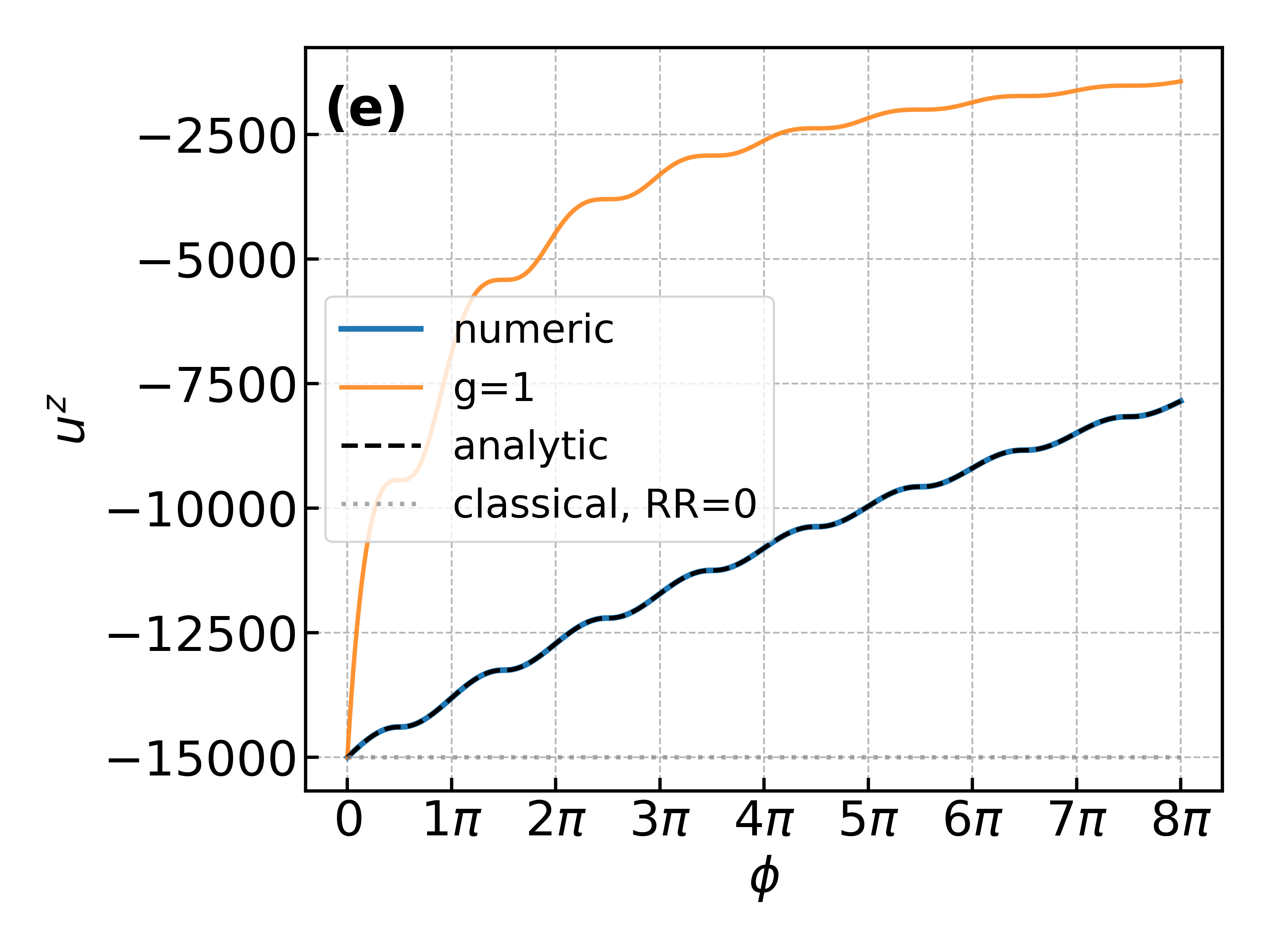}
    \includegraphics[width=0.48\linewidth]{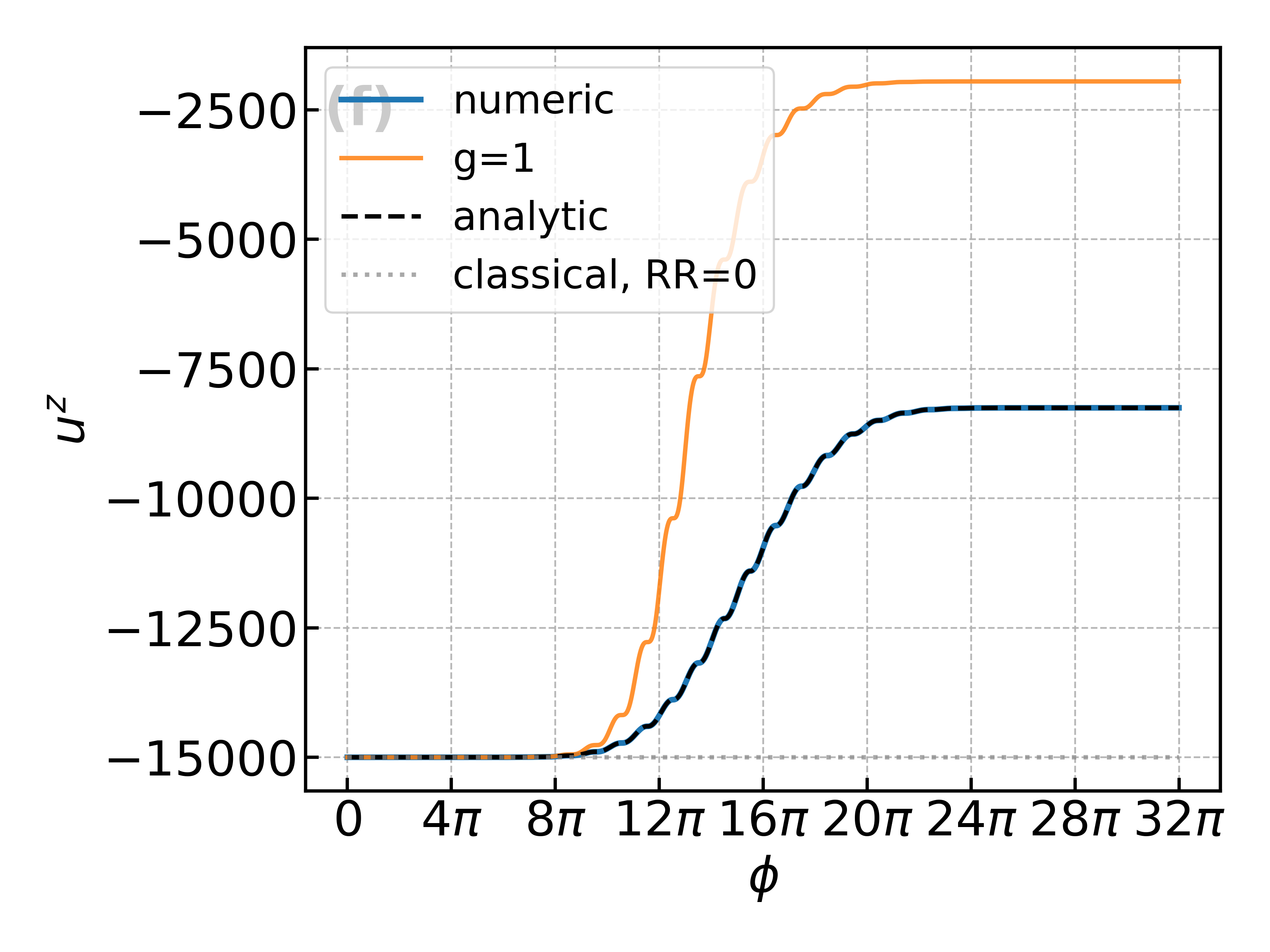}
    \caption{Evolution of the 4-velocity $u^\mu$ components in the classical RR case ($g = 1$) (orange) and the semiclassical Gaunt-modified case (blue) for (a) a monochromatic wave and (b) a finite-duration Gaussian pulse with $a_0=50$, $\gamma_0 = 1.5 \times 10^4$, $\omega\tau = 10$, $\phi_0 = 5\omega\tau$ and $\tau_R = 10^{-8}$. Dashed black line is analytical solution. Case of a counter-propagating relativistic particle.}
    \label{fig:four_velocity}
\end{figure}

Figure \ref{fig:four_velocity} shows the evolution of the four-velocity components during the interaction with the laser field. The initial four-velocity is chosen as
\[
u_0^\mu = (\gamma_0, 0, 0, -\gamma_0\beta),\qquad
\beta = \sqrt{1 - 1 / \gamma_0^2}
\]
with $\gamma_0 = 1.5 \times 10^4$. The particle therefore initially propagates along the negative $z$ direction, corresponding to a head-on collision geometry with the laser pulse.

Panels [Fig. \ref{fig:four_velocity}(a)] and [Fig. \ref{fig:four_velocity}(b)] show the temporal component $u^0$, which determines the particle energy. In the absence of radiation reaction, corresponding to the Lorentz-force solution, $u^0$ remains constant on average, consistent with the Lawson--Woodward theorem. When radiation reaction is included, the particle continuously loses energy through radiation emission, leading to a monotonic decrease of $u^0$. The strongest energy losses occur in the classical Landau--Lifshitz limit ($g=1$), whereas inclusion of the Gaunt factor $g(\chi)<1$ suppresses the radiation-reaction force and therefore reduces the rate of energy loss. For the monochromatic plane wave [Fig. \ref{fig:four_velocity}(a)], the decrease persists throughout the interaction. In contrast, for the finite Gaussian pulse [Fig. \ref{fig:four_velocity}(b)], the energy loss is localized within the main pulse envelope $\Delta\phi\sim \omega\tau$; after the field amplitude vanishes, $u^0$ approaches a constant asymptotic value, indicating that radiation emission has effectively ceased.

Panels [Fig. \ref{fig:four_velocity}(c)] and [Fig. \ref{fig:four_velocity}(d)] show the transverse component $u^x$. The oscillatory behavior is primarily driven by the transverse laser field and closely follows the Lorentz-force solution. Radiation reaction introduces only moderate corrections to the transverse dynamics, although a gradual reduction of the oscillation amplitude is visible in the pulsed case due to the combined effects of radiative damping and the finite pulse envelope. After the pulse has passed, the transverse oscillations disappear together with the external field.

Panels [Fig. \ref{fig:four_velocity}(e)] and [Fig. \ref{fig:four_velocity}(f)] present the longitudinal component $u^z$, which characterizes the particle motion along the initial propagation direction. Since the particle initially counter-propagates with respect to the laser wave, the quantity $u^z$ increases from its initial negative value as the particle decelerates under radiation losses. The increase is most pronounced in the classical Landau--Lifshitz case, while inclusion of the Gaunt factor leads to weaker deceleration and correspondingly smaller deviations from the Lorentz-force trajectory. For the Gaussian pulse [Fig. \ref{fig:four_velocity}(f)], $u^z$ tends toward a constant asymptotic value once the particle leaves the interaction region and the radiation-reaction force vanishes.

\section{Cycle-Averaged Dynamics and Poincar\'e Map}
\label{sec:averaged_dynamics}

\subsection{Slow-Flow Approximation and Effective Drift}

Quantum radiation reaction affects electron dynamics directly through the light-front momentum damping factor $h(\phi)$ and indirectly through the quantum suppression factor $g(\chi)$ (the Gaunt factor). The local quantum nonlinearity parameter is
\begin{equation}
\chi(\phi) = \frac{\chi_0(\phi)}{h(\phi)}, \qquad \chi_0(\phi) = \frac{\rho_0}{m}\sqrt{\sum_j\xi_j^2 \psi_j'^{2}(\phi)}.
\end{equation}
For a monochromatic plane wave with normalized vector potential amplitude $a_0$, the unsuppressed parameter reduces to $\chi_0(\phi) = \chi_{00}|\cos\phi|$, where
\begin{equation}
\chi_{00} = \frac{\rho_0 a_0}{m}.
\end{equation}

To capture quantum recoil accurately across both the semi-classical ($\chi \ll 1$) and strong-quantum ($\chi \gtrsim 1$) regimes, we employ the high-precision fitting formula for the Gaunt factor from Eq.~\eqref{eq:gaunt_fit}:
\begin{equation}\label{eq:gaunt_full}
g(\chi) = \left[ 1 + 18\chi + 69\chi^2 + 73\chi^3 + 5.806\chi^4 \right]^{-1/3}.
\end{equation}

The differential equation governing the light-front variable $h(\phi)$ is then
\begin{equation}\label{eq:h_exact_ode}
\frac{dh}{d\phi} = K_{\mathrm{eff}} \cos^2\phi \,
g\!\left( \frac{\chi_{00}|\cos\phi|}{h(\phi)} \right),
\end{equation}
where $K_{\mathrm{eff}} = \frac{2}{3} \alpha \eta \rho_0 a_0^2
= \tau_R \rho_0 a_0^2$ collects the classical radiation-reaction
prefactors, with $\alpha \approx 1/137$ being the fine-structure
constant, $\eta = \omega_0/m$ the normalized photon energy,
$\rho_0 = \gamma_0(1-\beta_{z,0})$ the initial light-front velocity
factor, and $\tau_R = \frac{2}{3}\alpha\eta$ the normalized classical
radiation parameter \cite{DiPiazza2008}. A summary of these physical
variables, mathematical definitions, and characteristic parameters is
provided in Table~\ref{tab:units}.

We define the cycle-averaged attenuation function as
\begin{equation}\label{eq:G_integral}
\mathcal{G}(\chi) \equiv \frac{2}{\pi}
\int_0^{\pi/2} \cos^2\phi \, g(\chi\cos\phi)\,d\phi.
\end{equation}
The validity of the cycle-averaging approximation requires the relative change of $h$ over one optical period to remain small, as quantified by the criterion in Eq.~\eqref{eq:validity_criterion}.
Consequently, $h(\phi)$ evolves on a phase scale much longer than one
optical period $2\pi$. Freezing $h(\phi) \simeq h_n \equiv h(2\pi n)$
over the $n$-th cycle $\phi \in [2\pi n,2\pi(n+1)]$ yields the
cycle-averaged slow-flow equation
\begin{equation}\label{eq:h_slowflow}
\frac{dh}{d\phi} \simeq K_{\mathrm{eff}} \,
\mathcal{G}(\chi_n), \qquad
\chi_n = \frac{\chi_{00}}{h_n}.
\end{equation}

\begin{figure}[t]
    \centering
    \includegraphics[width=0.5\linewidth]{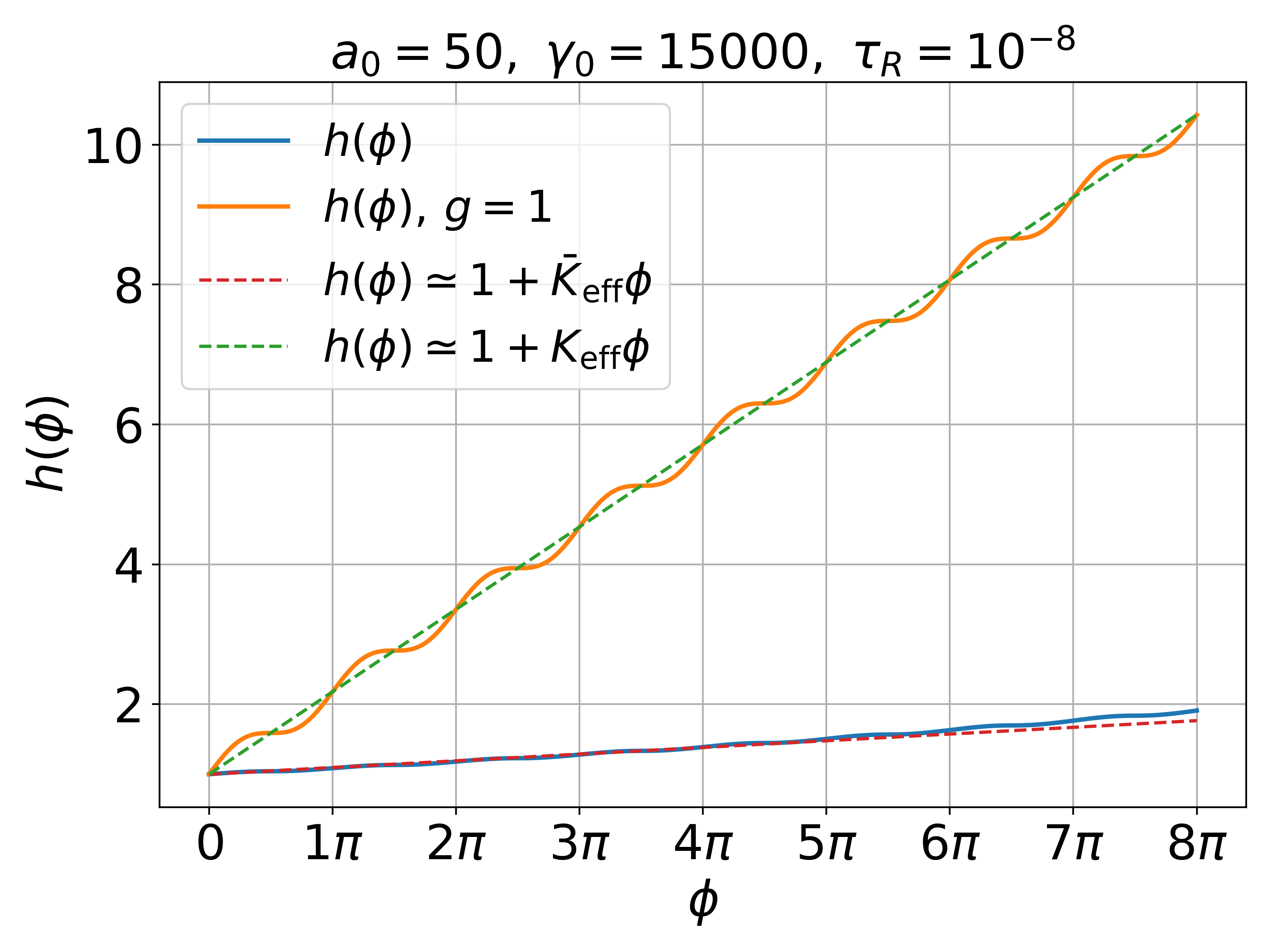}
    \caption{Evolution of the light-front damping factor $h(\phi)$ for a monochromatic plane wave in a counter-propagating relativistic electron--laser configuration. Results are shown for fixed parameters $a_0 = 50$, $\gamma_0 = 1.5 \times 10^4$, and $\tau_R = 10^{-8}$. The classical radiation-reaction case ($g=1$, orange) exhibits linear growth with slope $K_{\mathrm{eff}}/2$, while the semiclassical Gaunt-corrected case (blue) shows a reduced effective drift $\bar{K}_{\mathrm{eff}} = K_{\mathrm{eff}}\mathcal{G}(\chi_n)$, reflecting quantum suppression of radiative energy loss.}
    \label{fig:h_linear}
\end{figure}

Over phase intervals where $h_n$ varies slowly, the effective drift rate $\bar{K}_{\mathrm{eff}} = K_{\mathrm{eff}}\mathcal{G}(\chi_n)$ remains nearly constant, producing a local quasi-linear growth $h(\phi) \approx 1 + \bar{K}_{\mathrm{eff}}\phi$. Figure~\ref{fig:h_linear} illustrates this cycle-averaged slow-flow evolution. Compared to classical radiation reaction ($g=1$), where the average slope is $K_{\mathrm{eff}}/2$, the Gaunt factor reduces the effective growth rate to $\bar{K}_{\mathrm{eff}}$, resulting in a noticeably slower growth of $h(\phi)$.

\subsubsection{Asymptotic Limits and Physical Properties}

While Eq.~\eqref{eq:G_integral} is evaluated numerically via adaptive Gauss-Legendre quadrature in the dynamical solver, its behavior is fully characterized by its asymptotic limits:
\begin{enumerate}
    \item \textbf{Classical / Weak-Quantum Limit ($\chi_n \ll 1$):} Expanding $g(\chi) = 1 - 6\chi + 49\chi^2 + \mathcal{O}(\chi^3)$ yields
    \begin{equation}\label{eq:G_weak}
    \mathcal{G}(\chi_n) = \frac{1}{2} - \frac{8}{\pi}\chi_n + \frac{147}{8}\chi_n^2 + \mathcal{O}(\chi_n^3).
    \end{equation}
    In the limit $\chi_n \to 0$, $\mathcal{G}(0) = 1/2$, recovering the classical cycle-averaged drift $dh/d\phi = K_{\mathrm{eff}}/2$.
    
    \item \textbf{Deep-Quantum Limit ($\chi_n \gg 1$):} Using the leading asymptotic power law $g(\chi) \approx 0.556 \, \chi^{-4/3}$, explicit integration gives
    \begin{equation}\label{eq:G_strong}
    \mathcal{G}(\chi_n) \approx 0.397 \, \chi_n^{-4/3}.
    \end{equation}
    This power law reflects the severe quantum suppression of radiation reaction in extreme electromagnetic fields.
\end{enumerate}

The physical feedback loop operates as follows: as radiation reaction progresses, $h(\phi)$ grows monotonically ($dh/d\phi > 0$). Because $\chi_n = \chi_{00}/h_n$, an increasing $h_n$ suppresses the effective quantum parameter $\chi_n$. Consequently, the local Gaunt factor $g(\chi)$ increases back toward unity. Thus, quantum suppression gradually weakens as the electron decelerates, and the effective drift $\bar{K}_{\mathrm{eff}} = K_{\mathrm{eff}}\mathcal{G}(\chi_n)$ approaches its classical value $K_{\mathrm{eff}}/2$.

\subsection{Discrete Stroboscopic Poincar\'e Map and Stability Analysis}
\label{subsec:poincare_map}

The slow-flow continuous dynamics may be equivalently cast as a discrete stroboscopic Poincar\'e map sampled once per optical period \cite{SagdeevZaslavsky1988,SarachikSchappert1970}. Defining the discrete phase sampling points $\phi_n = 2\pi n$ with $h_n \equiv h(\phi_n)$, integration of the exact evolution equation over one period yields
\begin{equation}
h_{n+1} - h_n = \int_{\phi_n}^{\phi_n+2\pi} \frac{dh}{d\phi} \, d\phi.
\end{equation}
Under the cycle-averaged adiabatic approximation, this generates the one-dimensional map
\begin{equation}\label{eq:poincare_map}
h_{n+1} = f(h_n) \equiv h_n + 2\pi K_{\mathrm{eff}} \, \mathcal{G}\!\left( \frac{\chi_{00}}{h_n} \right),
\end{equation}
where $K_{\mathrm{eff}} = \tau_0 \rho_0 a_0^2$ collects the effective coupling parameters and $\mathcal{G}(\chi_n)$ is the cycle-averaged attenuation integral defined in Eq.~\eqref{eq:G_integral}.

The applicability of the adiabatic cycle-averaging approximation relies on the relative growth per optical period remaining sufficiently small:
\begin{equation}\label{eq:validity_criterion}
\frac{\Delta h_n}{h_n} = \frac{2\pi K_{\mathrm{eff}} \, \mathcal{G}(\chi_n)}{h_n} \ll 1.
\end{equation}
For standard interaction parameters (see Table \ref{tab:units}), quantum suppression restricts the initial relative increment per cycle to $\Delta h_1 / h_1 \approx 0.166$. As $h_n$ grows over subsequent cycles, $\Delta h_n / h_n$ falls rapidly below $0.03$. For lower field intensities or weaker radiation-reaction coupling ($a_0 = 100$, $\tau_R = 1.0 \times 10^{-9}$), $\Delta h_1 / h_1 \approx 0.025$.

To benchmark the precision of the stroboscopic map, we track the relative error against direct numerical integration of the un-averaged differential equation \eqref{eq:h_exact_ode}:
\begin{equation}
\delta_n = \frac{|h_n^{\mathrm{exact}} - h_n^{\mathrm{map}}|}{h_n^{\mathrm{exact}}}.
\end{equation}
For $a_0 = 50$, the relative error remains bounded below $\delta_n \lesssim 9.5\%$ throughout the fast transient phase and settles to $\delta_{100} \approx 1.6\%$ at $n = 100$ optical cycles. For $a_0 = 100$, the cumulative relative error remains strictly bounded below $\delta_n < 0.5\%$ across all cycles, demonstrating that the discrete map accurately reproduces the continuous phase-space dynamics while offering significant computational savings.

\subsubsection{Dynamical Stability and Feedback Dynamics}

The stability and non-chaotic nature of the iteration $h_{n+1} = f(h_n)$ can be demonstrated by examining the map derivative $f'(h)$. Differentiating Eq.~\eqref{eq:poincare_map} with respect to $h_n$ gives
\begin{equation}\label{eq:map_derivative}
f'(h_n) = 1 - 2\pi K_{\mathrm{eff}} \frac{\chi_{00}}{h_n^2} \mathcal{G}'(\chi_n).
\end{equation}
Because the Gaunt factor $g(\chi)$ is a strictly monotonically decreasing function of $\chi$, the cycle-averaged attenuation $\mathcal{G}(\chi_n)$ likewise satisfies $\mathcal{G}'(\chi_n) < 0$ for all $\chi_n > 0$. Consequently, the second term in Eq.~\eqref{eq:map_derivative} is strictly positive, yielding
\begin{equation}
f'(h_n) > 1 \qquad \forall \, h_n > 0.
\end{equation}
Since $f'(h_n)$ is strictly positive and monotonic, the map $f$ is a 1D orientation-preserving diffeomorphism without critical points or folding mechanisms. This rigorously precludes period-doubling bifurcations, subharmonic resonances, or deterministic chaos.

\begin{figure}[t]
    \centering
    \includegraphics[width=0.48\linewidth]{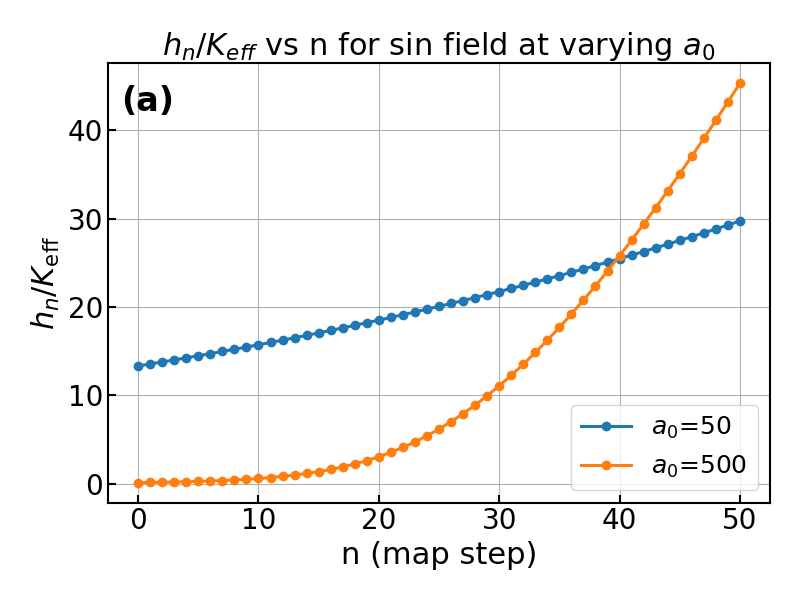}
    \includegraphics[width=0.48\linewidth]{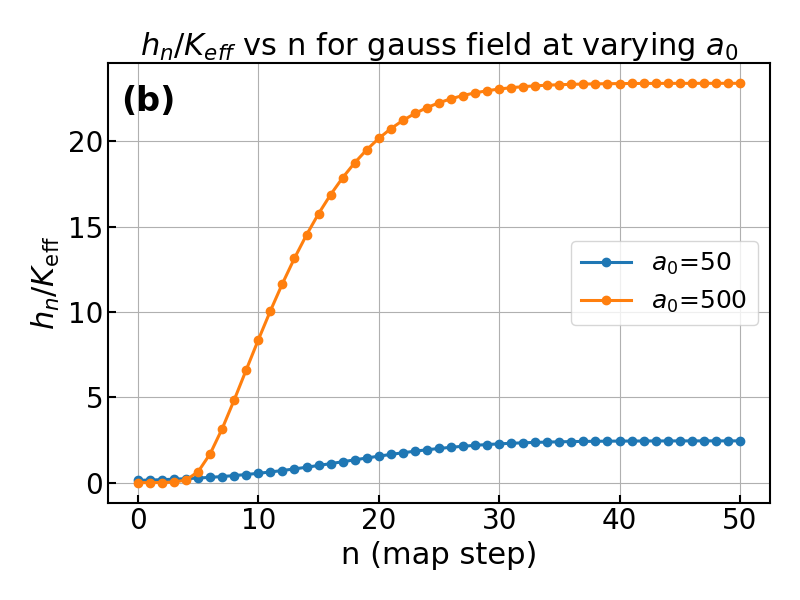}
    \caption{Cycle-to-cycle evolution of the normalized light-front damping factor $h_n / K_{\mathrm{eff}}$ as a function of cycle number $n$ for different laser amplitudes $a_0$. Panels (a) and (b) correspond to a monochromatic plane wave and a finite-duration Gaussian pulse, respectively. Normalization by $K_{\mathrm{eff}} = \tau_0 \rho_0 a_0^2$ collapses all curves toward the universal classical linear slope of $\pi$, while early-cycle departures highlight quantum suppression for larger $a_0$. Curves correspond to $a_0 = 50$ (peak $\chi_0 \sim 4$) and $a_0 = 500$ (peak $\chi_0 \sim 40$), with fixed initial parameters $\gamma_0 = 1.5 \times 10^4$, $\omega\tau = 10$, $\tau_R = 10^{-9}$.}
    \label{fig:hn_map}
\end{figure}

The discrete map $h_{n+1} = f(h_n)$ illuminates the transition across interaction regimes when expressed in terms of the rescaled variable $h_n / K_{\mathrm{eff}}$:
\begin{enumerate}
    \item \textbf{Affine Classical Limit ($\chi_n \ll 1$):} As $\chi_n \to 0$, $\mathcal{G}(\chi_n) \to 1/2$, and Eq.~\eqref{eq:poincare_map} reduces to the normalized affine map:
    \begin{equation}
    \frac{h_{n+1}}{K_{\mathrm{eff}}} \approx \frac{h_n}{K_{\mathrm{eff}}} + \pi.
    \end{equation}
    This establishes a universal asymptotic trajectory slope $\Delta (h_n / K_{\mathrm{eff}}) / \Delta n \to \pi \approx 3.14$, shared across all values of $a_0$ in the late-stage regime shown in Fig.~\ref{fig:hn_map}(b).
    
    \item \textbf{Cross-Regime Trajectory:} As described in Sec.~\ref{sec:averaged_dynamics}, the feedback mechanism ($\chi_n \propto 1/h_n$) ensures a smooth transition from quantum-suppressed drift rate $\bar{K}_{\mathrm{eff}}(h_n)$ at early times ($\chi_n \gtrsim 1$) to classical affine growth at late times ($\chi_n \ll 1$).
\end{enumerate}

Because $h_n$ relates to the light-front momentum via $n \cdot u(\phi_n) = \rho_0 / h_n$, linear growth $h_n \propto n$ in the classical limit corresponds to an inverse-linear decay of light-front momentum.

\section{Conclusion}

In this work we have shown that the Landau--Lifshitz equation with semiclassical Gaunt-factor correction remains integrable in a plane-wave background. Exploiting the fact that, in a plane wave, the quantum nonlinearity parameter $\chi$ depends solely on the light-front momentum $n \cdot u$, the full 4D differential dynamics reduces to a single scalar quadrature governing energy dissipation $h(\phi)$. The four-velocity $u^\mu(\phi)$ and trajectory $x^\mu(\phi)$ then follow in closed algebraic form, with the classical Di~Piazza solution \cite{DiPiazza2008} recovered in the limit $g(\chi)\to1$.

This deterministic light-front reduction complements recent kinetic and stochastic analytical frameworks (e.g., Blackburn \cite{Blackburn2024}), which describe beam energy variance, higher-order distribution moments, and photon straggling in strong fields. While kinetic models predict ensemble spectrum broadening, our deterministic single-particle solution provides an exact analytical baseline against which continuous radiation-reaction modules in QED-PIC codes (such as \textsc{Smilei} and \textsc{Epoch}) can be validated.

The Gaunt-factor correction captures the deterministic reduction of radiative losses due to quantum recoil while preserving a classical equation-of-motion structure. However, it does not include inherently quantum stochastic effects such as random photon emission and discrete recoil events. Such stochastic features become essential in the regime $\chi \gtrsim 1$, where emission becomes strongly probabilistic and cascade-like dynamics may emerge in both laboratory and astrophysical environments.

By reducing the semiclassical dynamics to a simple 1D quadrature, our work establishes a direct analytical bridge between classical integrable radiation reaction and quantum-corrected model equations. It provides a controlled baseline against which both deterministic Gaunt-factor approaches and stochastic QED simulations can be compared, clarifying the domain of validity of reduced models in high-field physics.

Future work may extend this framework to include stochastic emission processes on top of the exact semiclassical trajectory, or explore plane-wave structures with nontrivial polarization and envelope dynamics.

\bibliographystyle{apsrev4-2}
\bibliography{bibliography}

@article{DiPiazza2012,
  author  = {Di Piazza, A. and M{\"u}ller, C. and Hatsagortsyan, K. Z. and Keitel, C. H.},
  title   = {Extremely high-intensity laser interactions with fundamental quantum systems},
  journal = {Reviews of Modern Physics},
  volume  = {84},
  pages   = {1177--1228},
  year    = {2012},
  doi     = {10.1103/RevModPhys.84.1177}
}

@article{MarklundShukla2006,
  author  = {Marklund, M. and Shukla, P. K.},
  title   = {Nonlinear collective effects in photon-photon and photon-plasma interactions},
  journal = {Reviews of Modern Physics},
  volume  = {78},
  pages   = {591--640},
  year    = {2006},
  doi     = {10.1103/RevModPhys.78.591}
}

@article{Gonoskov2022,
  author  = {Gonoskov, A. and Blackburn, T. G. and Marklund, M. and Bulanov, S. S.},
  title   = {Charged particle motion and radiation in strong electromagnetic fields},
  journal = {Reviews of Modern Physics},
  volume  = {94},
  number  = {4},
  pages   = {045001},
  year    = {2022},
  doi     = {10.1103/RevModPhys.94.045001}
}

@article{DiPiazza2008,
  author  = {Di Piazza, A.},
  title   = {Exact Solution of the Landau-Lifshitz Equation in a Plane Wave},
  journal = {Letters in Mathematical Physics},
  volume  = {83},
  number  = {3},
  pages   = {305--313},
  year    = {2008},
  doi     = {10.1007/s11005-008-0228-9}
}

@article{Bulanov2011,
  author  = {Bulanov, Sergei V. and Esirkepov, Timur Zh. and Kando, Masaki and Koga, James K. and Bulanov, Stepan S.},
  title   = {Lorentz-Abraham-Dirac versus Landau-Lifshitz radiation friction force in the ultrarelativistic electron interaction with electromagnetic wave (exact solutions)},
  journal = {Phys. Rev. E},
  volume  = {84},
  number  = {5},
  pages   = {056605},
  year    = {2011},
  doi     = {10.1103/PhysRevE.84.056605}
}

@book{SokolovTernov1986,
  author    = {Sokolov, A. A. and Ternov, I. M.},
  title     = {Radiation from Relativistic Electrons},
  publisher = {American Institute of Physics},
  year      = {1986},
  address   = {New York}
}

@book{LandauLifshitz,
  author    = {Landau, L. D. and Lifshitz, E. M.},
  title     = {The Classical Theory of Fields},
  publisher = {Pergamon Press},
  year      = {1975},
  edition   = {4th}
}

@book{BaierKatkov1998,
  author    = {Baier, V. N. and Katkov, V. M. and Strakhovenko, V. M.},
  title     = {Electromagnetic Processes at High Energies in Oriented Single Crystals},
  publisher = {World Scientific},
  year      = {1998},
  address   = {Singapore}
}

@article{BaierKatkovStrakhovenko1974,
  author  = {Baier, V. N. and Katkov, V. M. and Strakhovenko, V. M.},
  title   = {The kinetics and integral characteristics of synchrotron radiation at high energies},
  journal = {Soviet Physics JETP},
  volume  = {39},
  number  = {1},
  pages   = {36--43},
  year    = {1974}
}

@article{BellKirk2008,
  author  = {Bell, A. R. and Kirk, J. G.},
  title   = {Possibility of prolific pair production with high-power lasers},
  journal = {Physical Review Letters},
  volume  = {101},
  pages   = {200403},
  year    = {2008},
  doi     = {10.1103/PhysRevLett.101.200403}
}

@article{Elkina2011,
  author  = {Elkina, N. V. and others},
  title   = {QED cascades induced by circularly polarized laser fields},
  journal = {Physical Review Special Topics - Accelerators and Beams},
  volume  = {14},
  pages   = {054401},
  year    = {2011},
  doi     = {10.1103/PhysRevSTAB.14.054401}
}

@article{Ridgers2014,
  author  = {Ridgers, C. P. and Kirk, J. G. and Duclous, R. and
             Blackburn, T. G. and Brady, C. S. and Bennett, K. and
             Arber, T. D. and Bell, A. R.},
  title   = {Modelling gamma-ray photon emission and pair production
             in high-intensity laser-matter interactions},
  journal = {Journal of Computational Physics},
  volume  = {260},
  pages   = {273--285},
  year    = {2014},
  doi     = {10.1016/j.jcp.2013.12.007}
}

@article{Bulanov2010,
  author  = {Bulanov, S. S. and Mur, V. D. and Narozhny, N. B. and Nees, J. and Popov, V. S.},
  title   = {Multiple Colliding Electromagnetic Pulses: A Way to Lower the Threshold of Pair Production from Vacuum},
  journal = {Physical Review Letters},
  volume  = {104},
  number  = {22},
  pages   = {220404},
  year    = {2010},
  doi     = {10.1103/PhysRevLett.104.220404}
}

@article{Bulanov2024,
  author   = {Bulanov, S. V. and Grittani, G. M. and Shaisultanov, R. and Esirkepov, T. Zh. and Ridgers, C. P. and Bulanov, S. S. and Russell, B. K. and Thomas, A. G. R.},
  title    = {On the energy spectrum evolution of electrons undergoing radiation cooling},
  journal  = {Fundamental Plasma Physics},
  volume   = {9},
  pages    = {100036},
  year     = {2024},
  doi      = {10.1016/j.fpp.2024.100036}
}

@article{Esirkepov2015,
title = {Attractors and chaos of electron dynamics in electromagnetic standing waves},
journal = {Physics Letters A},
volume = {379},
number = {36},
pages = {2044-2054},
year = {2015},
issn = {0375-9601},
doi = {https://doi.org/10.1016/j.physleta.2015.06.017},
url = {https://www.sciencedirect.com/science/article/pii/S0375960115005368},
author = {Timur Zh. Esirkepov and Stepan S. Bulanov and James K. Koga and Masaki Kando and Kiminori Kondo and Nikolay N. Rosanov and Georg Korn and Sergei V. Bulanov}
}

@article{Narozhny2015,
  author  = {Narozhny, N. B. and Fedotov, A. M.},
  title   = {Extreme light physics},
  journal = {Contemporary Physics},
  volume  = {56},
  number  = {3},
  pages   = {249--268},
  year    = {2015},
  doi     = {10.1080/00107514.2015.1028768}
}

@article{Fedotov2017,
  author  = {Fedotov, A. M.},
  title   = {Conjecture of perturbative QED breakdown at $\alpha\chi^{2/3} \gtrsim 1$},
  journal = {Journal of Physics: Conference Series},
  volume  = {826},
  number  = {1},
  pages   = {012027},
  year    = {2017},
  doi     = {10.1088/1742-6596/826/1/012027}
}

@article{Niel2018,
  author  = {Niel, F. and Riconda, C. and Amiranoff, F. and
             Duclous, R. and Grech, M.},
  title   = {From quantum to classical modeling of radiation reaction:
             A focus on stochasticity effects},
  journal = {Physical Review E},
  volume  = {97},
  number  = {4},
  pages   = {043209},
  year    = {2018},
  doi     = {10.1103/PhysRevE.97.043209}
}

@article{Niel2018PPCF,
  title = {From quantum to classical modelling of radiation reaction: a focus on the radiation spectrum},
  author = {Niel, F. and Riconda, C. and Amiranoff, F. and Lobet, M. and Derouillat, J. and P{\'e}rez, F. and Vinci, T. and Grech, M.},
  journal = {Plasma Physics and Controlled Fusion},
  volume = {60},
  number = {9},
  pages = {094002},
  year = {2018},
  doi = {10.1088/1361-6587/aace22}
}

@article{Blackburn2024,
  author  = {Blackburn, T. G.},
  title   = {Analytical solutions for quantum radiation reaction in high-intensity lasers},
  journal = {Physical Review A},
  volume  = {109},
  number  = {2},
  pages   = {022234},
  year    = {2024},
  doi     = {10.1103/PhysRevA.109.022234}
}

@article{Ritus1985,
  author  = {Ritus, V. I.},
  title   = {Quantum effects of the interaction of elementary particles with an intense electromagnetic field},
  journal = {Journal of Soviet Laser Research},
  volume  = {6},
  pages   = {497--617},
  year    = {1985}
}

@book{SagdeevZaslavsky1988,
  author    = {Sagdeev, R. Z. and Usikov, D. A. and Zaslavsky, G. M.},
  title     = {Nonlinear Physics: From Pendulum to Turbulence and Chaos},
  publisher = {Harwood Academic Publishers},
  address   = {New York},
  year      = {1988}
}

@article{SarachikSchappert1970,
  author  = {Sarachik, E. S. and Schappert, G. T.},
  title   = {Classical theory of the scattering of intense laser radiation by free electrons},
  journal = {Physical Review D},
  volume  = {1},
  pages   = {2738--2753},
  year    = {1970},
  doi     = {10.1103/PhysRevD.1.2738}
}

\end{document}